\documentstyle[epsf,11pt]{article}
\begin{document}
\vskip -6.0cm
\noindent
\rightline{\bf KEK Preprint 97-252}
\rightline{\bf BELLE Preprint 98-1}
\rightline{\bf SAGA-HEP 127}
\renewcommand{\thefootnote}{\fnsymbol{footnote}}
  \begin{center}
\vskip 1.5cm
\begin{Large}
\centerline {\bf The  monitoring system for the aerogel Cherenkov counter}
\centerline {\bf  of the BELLE  detector }
\footnote{to be published in Nucl. Instrum. Meth. \bf A }
\end{Large}
\vskip 1cm
M.H.R Khan$^a$,
A. Murakami$^a$\footnote{Corresponding author. E-mail address: murakami@cc.saga-u.ac.jp},
T. Sumiyoshi$^b$,
T. Kuniya$^a$,
I. Adachi$^b$,\\
\vskip 0cm
R. Enomoto$^b$,
H. Hattori$^c$,
T. Iijima$^b$,
K. Kaneda$^d$, 
R. Kawabata$^c$,
\vskip 0cm
T. Ooba$^e$,
R. Suda$^f$,
K. Suzuki$^c$
and M. Watanabe$^e$\\
\vskip 1cm
{\it
$^a$Department of Physics, Saga University, Saga 840-8502, Japan\\
$^b$Physics Division, High Energy Accelerator Research Organization (KEK), Ibaraki 305-0801, Japan\\
$^c$Department of Physics, Chiba University, Chiba 260-8522, Japan\\
$^d$Applied Fiber Optics Section, Fujikura Limitted, Sakura, Chiba 285-0812, Japan\\
$^e$Department of Physics, Chuo University, Tokyo 112-8551, Japan\\
$^f$Department of Physics, Tokyo Metropolitan University, Tokyo 192-0397, Japan\\ }
\end{center}
\vskip 2.0cm
\begin{center}{\bf Abstract}  \end{center}                     
        We report on  a design and performances  of a   monitoring  system 
        developed for   the aerogel
        Cherenkov counters (ACC) of the  BELLE detector. The system  consists 
        of blue LEDs,  a diffuser box,  and  optical distributors  which 
        distribute the LED light to
        the ACC modules. The employed LED (NSPB series) has been observed to
        have high reliability on the long term stability and the temperature
        dependence.
        The diffuser box is employed to reduce the intrinsic
        non-uniformity of the
        LED light intensity . The overall performances of the present
        monitoring system on  uniformity and  intensity of the light output
        have been found 
        to  satisfy all the requirements
        for the monitoring.
         \par
         \vspace{0.3 in}     
                 
\newpage
\section{ Introduction}
\begin{figure}[p]
\vspace {6.0in}
\hspace{-1.0in} 
  \includegraphics{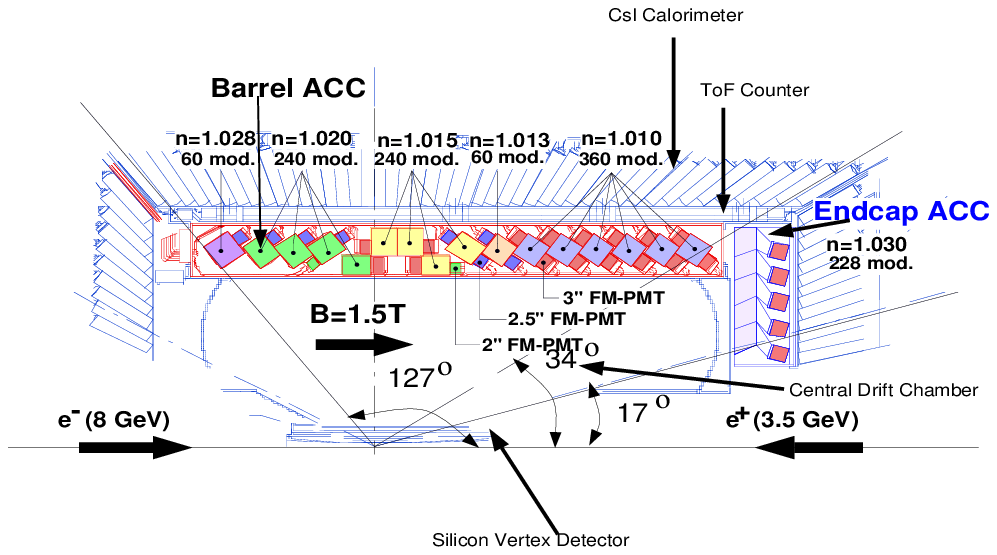}
\vspace{-3in}
\caption{  }
\label{fig:side_view}
\end{figure}

                  \hspace{0.2in}   
                 Particle identification plays an important role in CP 
                 violation studies in B-factory experiments. 
                 A threshold  aerogel Cherenkov counter (ACC) system will be 
                 used in the BELLE
                 experiment at the KEK B-factory to separate charged pions 
                 and kaons in the 
                 momentum 
                 range from 0.8 to 3.5 GeV/c [1,2,3] .  A side view 
                 of the  aerogel Cherenkov counter of the BELLE detector is
                 shown in Figure \ref{fig:side_view}. The details of the 
                 production method of silica aerogel for the BELLE experiment
                 are described in ref.[4].  \par 

\begin{figure}[p]
\vspace {6.5in}
\hspace{-1.0in} 
   \includegraphics{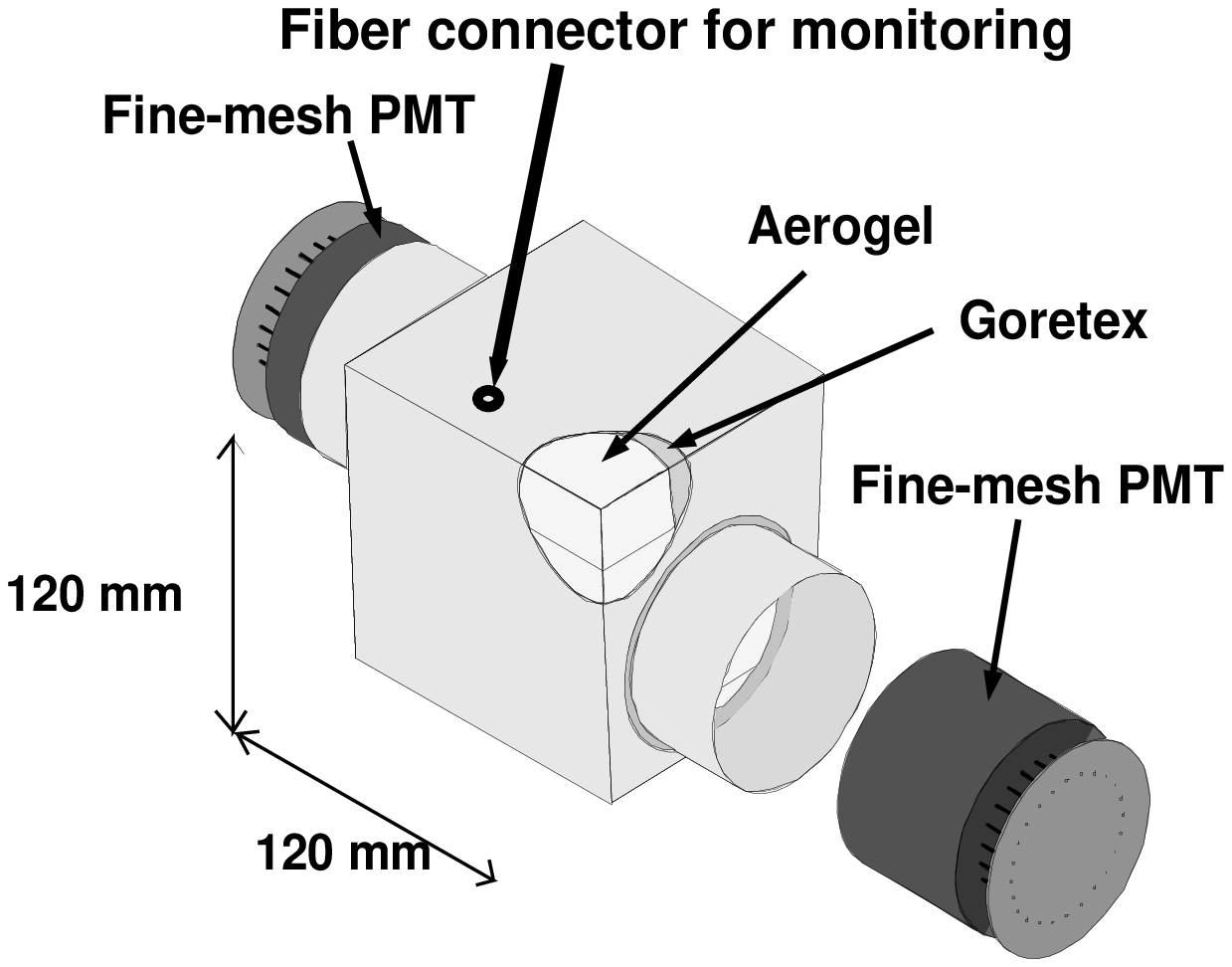}
\vspace{-3in}
  \caption{  }
\label{fig:5single_module}
\end{figure}

                 The ACC consists of 
                 1188 modules  and is equipped with a total of 1788 
                 fine-mesh (FM) PMTs to readout the produced Cherenkov 
                 photons [5]. A typical single module of the ACC is shown in 
                 Figure \ref{fig:5single_module}.   The FM-PMT
                 is one type of a photo-detector which can be operated in the
                 strong magnetic field [6]. The gain of the FM-PMT and light
                 yield of aerogels
                 may change 
                 during the 
                 period of data taking . In order to achieve high quality  
                 particle identification it is   necessary to monitor
                 the gain
                 of  PMTs and the light yield of aerogels of the ACC with 
                 a precision of a few percent
                 level.
                 \par
                 Since monitoring of the output signals from the PMTs of the
                 ACC 
                 does not distinguish the change in the gain of the PMTs from
                 that in the light yield of aerogels, we call the monitoring
                 of the PMT gain and the aerogel light yield simply the 
                 monitoring of the ACC. 
                 The  monitoring of these ACC modules  can be 
                 performed by distributing light pulses 
                 from one common light source through optical fibers or by 
                 equipping 
                 every counter with its own reference light pulser. The former
                 has some advantages over  the latter. In the former case we 
                 can 
                 easily control and monitor the light source itself. To 
                 build up
                 such a  monitoring system we need a powerful light 
                 source and a good  light distributing network  which can 
                 distribute
                 light pulses of almost equal intensity to all of the ACC
                 modules. \par
 
                 The  spectrum and timing of the monitoring light  
                 should be as similar as possible to that of the
                 Cherenkov light  generated by the 
                 incident particles.  Since the real signal has a few {\it ns}
                  rise
                 time, the monitoring light pulse should be as fast as 
                 possible.  In addition the light pulses should be  stable 
                  with respect to time and temperature enough to satisfy the 
                 requirements of the experiments. Some  monitoring systems  
                 for calorimeters  usually use a Xenon flash 
                  lamp[7]
                 , a light emitting diode [8] or a laser[9,10] as a light
                 source. The Xenon flash lamp generates
                  pulses of  
                 larger intensity but of wider  width  than a few hundred
                {\it ns} and the laser is not stable with respect to 
                 temperature
                 [10] . 
                 An LED is a much more  
                 convenient light source of  fast and stable pulses. As a light
                 source for the present ACC monitoring  system  we have 
                 chosen the  blue LED  of 
                 {NSPB series\footnotemark}
                 \footnotetext{The LED NSPB series is manufactured by NICHIA 
                 CHEMICAL INDUSTRIES, Ltd., Anan-shi, Tokushima 774, Japan.} 
                 [11] which has fast response, high intensity, and 
                 a wavelength range  close to that of the 
                  Cherenkov light detected with PMTs.  \par 
                            
                 In the present paper we  describe the design,
                 characteristics 
                 and the performance of the system to  be used 
                 to  monitor
                 a large number of the ACC modules  of the BELLE detector.   
                
                 \section{ General description of the ACC monitoring system }
                  \hspace {0.2in}
                    To monitor the 
                  ACC the  monitoring system should  supply roughly
                 the same light intensity as that of the Cherenkov light 
                 generated by the pions to each ACC module,
                 i.e. roughly   18
                 photoelectrons for
                 each FM-PMT of the ACC module, according to the beam test 
                 results [1]. 
                  The monitoring light  intensity  should be stable
                 enough ( $\sim$ 99 \%\ ) with respect to time and 
                 temperature
                  to
                 measure the possible  drift of the PMT gain and the light 
                 yield of aerogels in a long period and that
                due to  temperature change. The light distributing network
                 should have small light loss and high uniformity 
                ( $\sim$ 95 \%\ ) to  supply sufficiently intense and 
                uniform light to all
                of the 1188 ACC modules. The definition of the stability and 
                the uniformity will be discussed later in Sec.3.
                In the following subsections
                we describe each essential component of the present 
                monitoring system.
               
\subsection{The light source}
     \hspace {0.2in}
              From the  study of the performance of a  Xe-flasher lamp and 
             several
             blue LEDs, the blue LED NSPB series  has been chosen for
             our monitoring system. All types of LEDs of this series
             have the same optical characteristics. The LEDs of this series 
             are made of GaN having a single quantum well  structure [12].
             Due to the optical properties of GaN the LED of NSPB series
             can emit
             much  higher  intensity than other  conventional type LEDs
             available in the market. The peak wavelength of 
             the emission spectrum of the LED of this series   is 470 {\it nm}
             with a FWHM of 20 {\it nm} . Its luminous
             output and luminous intensity are 3.0 {\it mW} and 1.0 {\it cd} ,
             respectively
             at the DC forward voltage of 3.6 volt.  The maximum pulse forward
             current of the LED of this series is 100 {\it mA} and it can be 
             operated
             with a pulse width up to 10 {\it ms} . It has been observed that
             this
             LED can 
             be oprerated with  pulses of the width at least down to  8 
             {\it ns}.
             The typical operating temperature is 
             $-20^{\circ}$C   
             $\sim$ $+80^{\circ}$C . The temperature coefficient of the LED
             of this series will be discussed later. The maximum duty ratio 
             of this type of
              LED is
              1/10. \par 
                
 \subsection{The diffuser box}     
\begin{figure}[p]
\vspace {8.0in}
\hspace{0.0in}
  \includegraphics{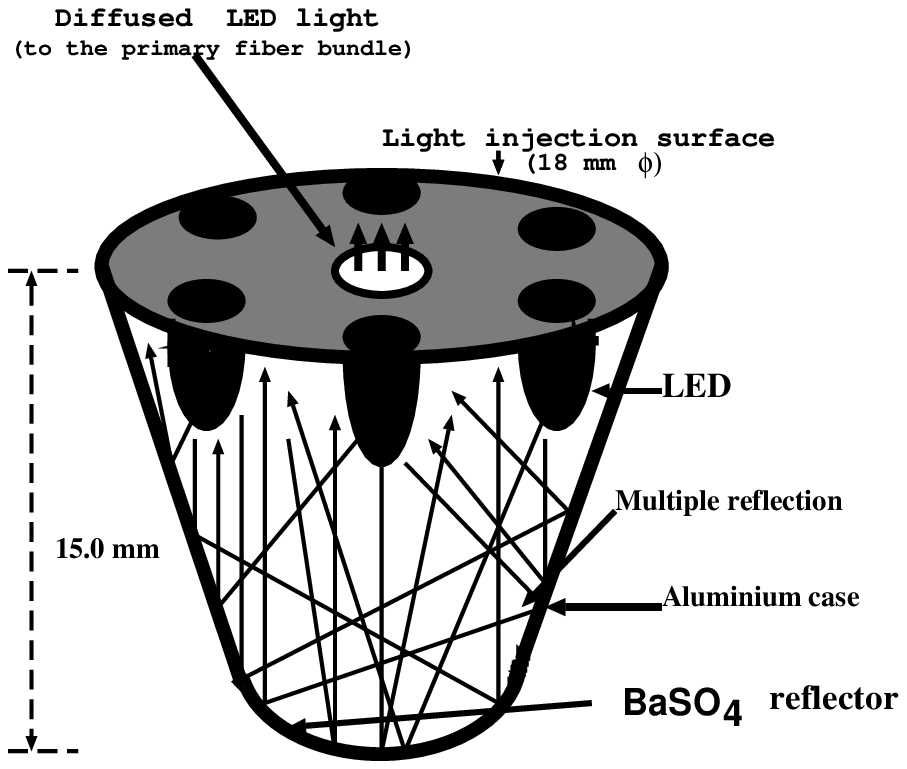}
\vspace{-4in}
\caption{  }
\label{fig:dbox}
\end{figure}

    \hspace {0.2in}
        The LED light is transmitted to a bundle of 19 pieces of 
        optical fibers (primary fiber bundle). However, when the LED light
        is directly injected into the fiber bundle, each of 19 fibers will
        not receive the same light yield because the  light intensity 
        distribution of the LED of
        NSPB series is not uniform in ${\theta}$ and  ${\phi}$ angle  due to 
        the non-symmetrical structure of the light emission part of the 
        LED [13] .   
        To reduce the intrinsic non-uniformity of the light intensity  on 
        the input surface
        of the primary fiber bundle, the LED light is diffused through 
        multiple 
        reflection  in a diffuser box. 
        To get uniformly diffused  LED light several 
        types of  light diffuser 
        boxes having different geometrical structure
        were studied. The  best one which has the highest light intensity 
        was selected. Figure \ref{fig:dbox}  shows the structural 
        view of the present
        diffuser box. It  is made of thin aluminum foil whose
        inner surface  is coated with the
        BaSO$_{4}$  reflector. The reflectivity of   BaSO$_{4}$  for the  
         wavelength of 470 {\it nm} was measured to be roughly  98 \%\ .
         The  size of the diffuser box is 15 {\it mm}  in height
        having a light injection surface of 18 {\it mm} in diameter. Six LEDs 
        of 
        NSPB500S are 
        attached on the light injection  surface of the box as shown in the 
        figure. Emitted light from six LEDs is reflected and diffused by the
        reflector  BaSO$_{4}$ on the inner surface.
        The diffused  LED light is transmitted  to the
        primary fiber bundle through an out-let hole of 
        an area of 7.1 {\it mm}${^2}$ on the light injection  surface. The 
        primary
        fiber bundle is directly connected to the out-let hole via an 
        optical fiber connector. \par 

 \subsection{The Light distributor}
\begin{figure}[p]
\vspace {5.0in}
\hspace{-1.5in}
     \includegraphics{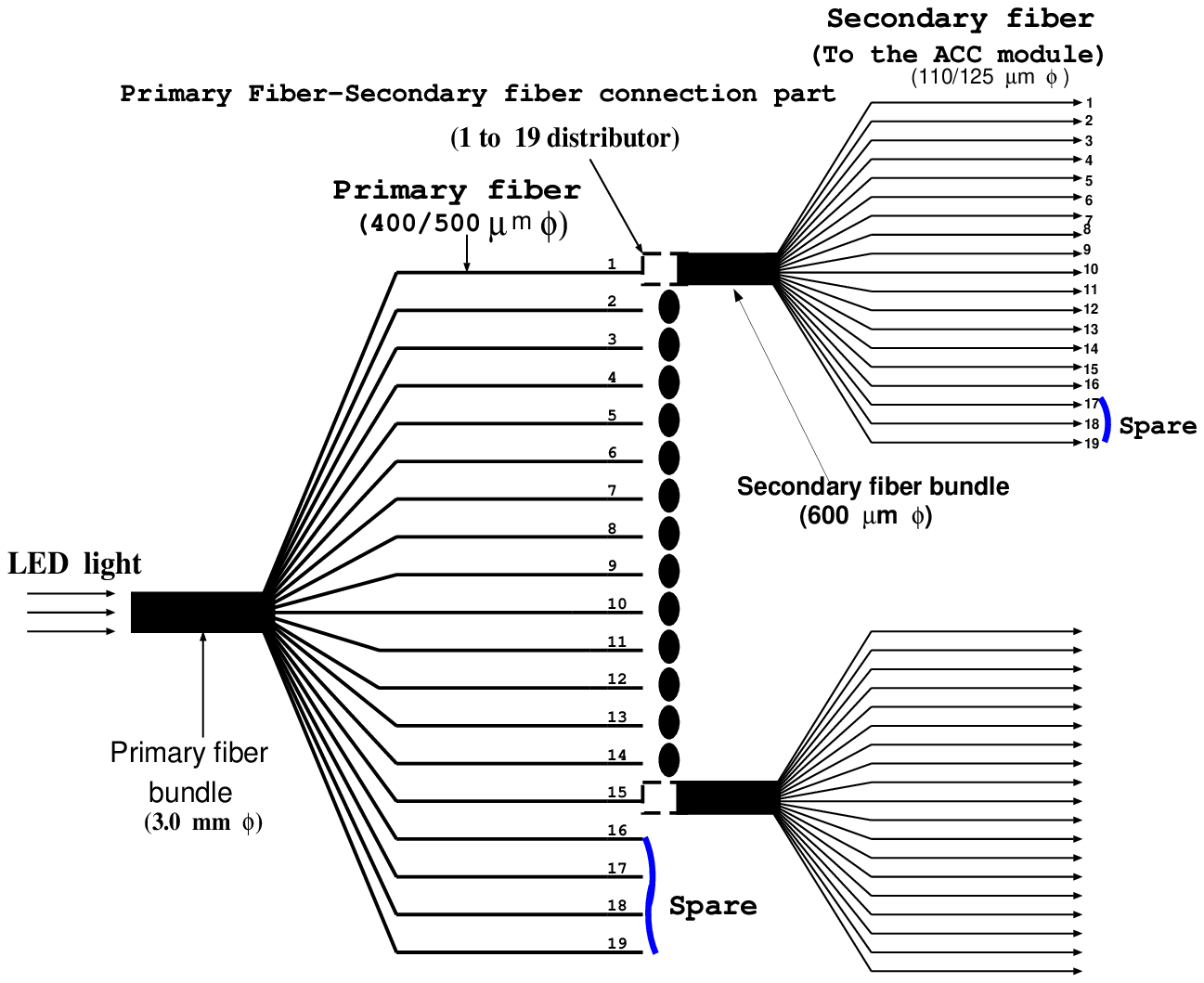}
\vspace{-2.5in}
\caption{  }
\label{fig:dist_set}
\end{figure}

   \hspace {0.2in}
        The  LED light should be distributed to all ACC modules uniformly with
        minimum light loss by an
        optical distributor network . 
        Figure \ref{fig:dist_set}  shows one set of the 
       {distributor\footnotemark}
                 \footnotetext{The light distributor network is manufactured 
                  by Fujikura Limitted, Sakura, Chiba 285, Japan}
        for the present monitoring system.  It consists of a primary fiber
        bundle having 19 primary quartz fibers.  
        Each of these primary fibers has  a length of 10 meters and a core/clad
       diameter of 400/500 $\mu m$. Fifteen primary fibers among 19 ones
       are  connected to 15 secondary fiber bundles as indicated in the 
       figure and the other 4 fibers are reserved for  spare.
       Each of the secondary  fiber bundles consists of 19 secondary 
       quartz fibers
       having a length of 0.5 $\sim$ 2.0 meter and a core/clad diameter of 
       110/125 $\mu m$. For each secondary fiber bundle the  
        16 secondary fibers are connected to 
       16  ACC modules one by one and other 3 fibers are for spare . \par

               The diameters of the primary fiber bundle and  the
        secondary  fiber bundle are about 3.0 {\it mm} and about
        600 $\mu m $ , respectively. 
        One set of the 
        distributing network can supply LED light pulses 
        to 240 ACC modules. Four sets of this network  monitor all  the
        960 ACC modules of the barrel part and another one set, 228 modules 
        of the endcap part of the ACC. For the barrel part 4 primary fiber
        bundles and 64 secondary
        fiber bundles are prepared; among which 4 secondary fiber bundles 
        are the spare ones. For the endcap part one primary fiber bundle
        and  17 secondary fiber bundles of the same type as for the
        barrel part 
         are in  fabrication  among which 2 secondary fiber bundles are
        the spare ones.
        Table \ref{tab:dist} shows  the statistics   of the
        distributor networks necessary to monitor all of the ACC modules
        of the BELLE detector. \par         
    
  \subsubsection{The mode scrambler} 
\begin{figure}[p]
\vspace {8.5in}
\hspace{-1.5in}
     \includegraphics{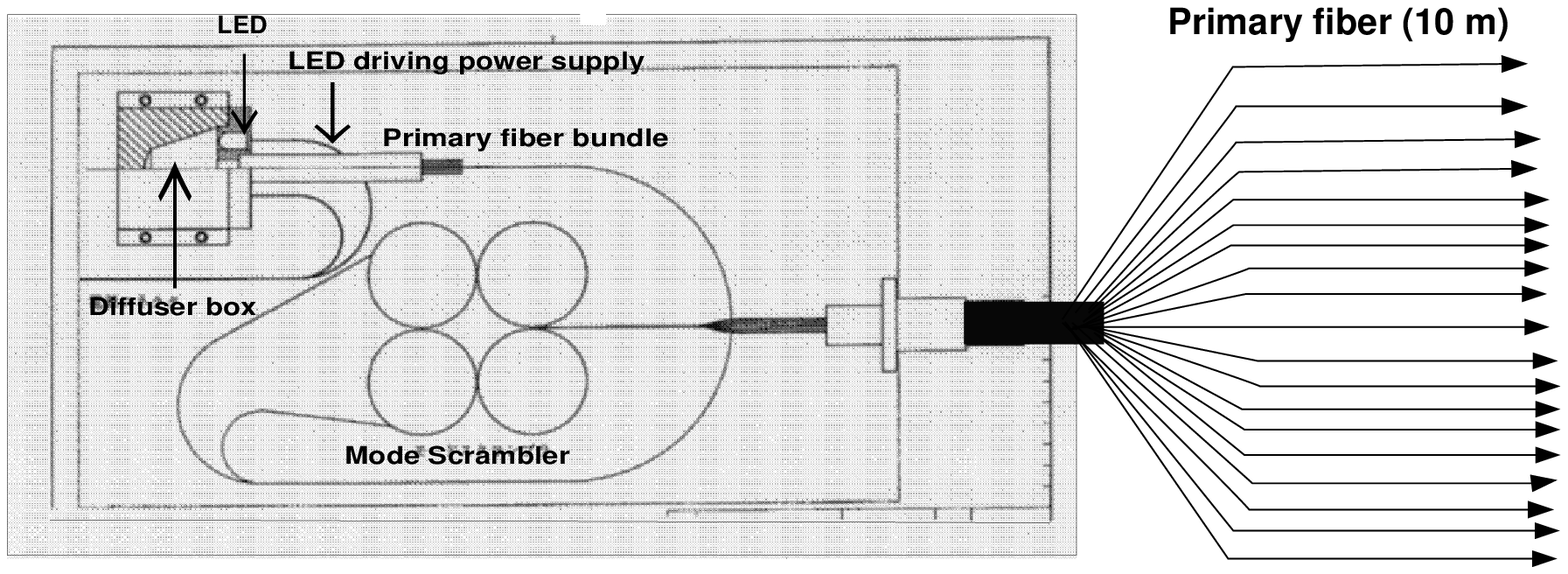}
\vspace{-5in} 
 \caption{   } 
 \label{fig:led_pf}
\end{figure}

     \hspace {0.2in}
        Figure  \ref{fig:led_pf}  shows the structural view of the
       connection part of the light diffuser box with 
       the primary fiber bundle. The  LED light accepted by the primary
       fiber bundle is 
       transported to the mode scrambler as shown in the figure. 
       The mode scrambler is the several round circular turns of each 
       primary fiber
       having a radius of 1.75 {\it cm}  which is the minimum bending radius 
      of the 10 {\it m} long primary fiber allowed to ensure no light loss in 
      the light transmission. 
       During  many revolutions of the light in the mode scrambler 
        the large off-axis angle component of the diffused  
        LED light is filtered out. As a result there is almost no
        light loss within the
        10 {\it m} long primary fiber when the bending curvature of the fiber
        at any point is larger than
        1.75 {\it cm}.    \par

  \subsubsection{ Connection part of the primary fiber with the
           secondary fiber bundle  }
\begin{figure}[p]
\vspace {6.0in}
\hspace{-0.5in}
      \includegraphics{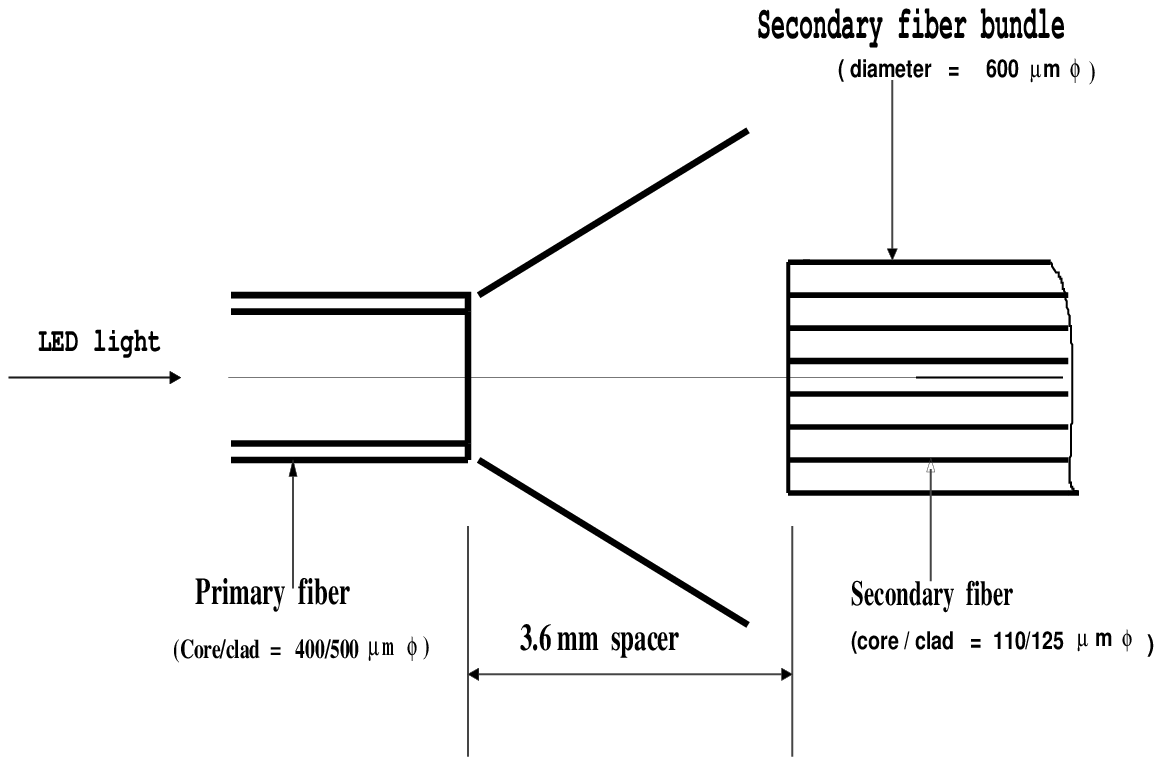}
\vspace{-4.5in}  
  \caption{  }
\label{fig:pf_sf1}
\end{figure}

    \hspace {0.2in}
         This connection part of the  monitoring system 
         affects  the uniformity 
        and the intensity of 
        the light transmitted . Thus this connection part is carefully designed
        and tested. The structural view of this connection part
        is  shown in Figure  \ref{fig:pf_sf1}. \par 
         
         The diameter of the secondary fiber bundle is greater than
        that  of each  primary fiber. To get  the uniform 
        LED light intensity on the input surface of the secondary fiber bundle
        sufficient space is necessary between the primary fiber and the 
        secondary fiber bundle.        
       The effect of the length of the space on the light intensity
        and uniformity was studied
        for  two  spacers  of 2.3 {\it mm} and 3.6 {\it mm} in length 
        between the primary fiber and the secondary
        fiber bundle.  It has been  found that the uniformity of light 
        intensity
        among the secondary
        fibers are about 95 \%\ and 89 \%\ for the 3.6 and 2.3 {\it mm} 
         spacing,
        respectively. 
        The light intensity for the  3.6 {\it mm} spacing
        is  $\sim$  50 \%\ of that for the   2.3 {\it mm} one.  
        Considering our requirements of light intensity and uniformity
        the 3.6 {\it mm} spacer  is employed between the primary fiber and the 
        secondary fiber bundle as shown in  Fig.6.   
         
 \section{Performance of the monitoring system}  
   \subsection{ Performance  of the  LED light source }
         \hspace {0.2in}     
                 The long term stability 
                 and temperature dependence of the LED  of NSPB  series
                 have been studied and are discussed in the following 
                 subsections. 
                 \par
             \subsubsection{Experimental arrangement  and data taking}
             \hspace {0.2in}
            Three pieces of LEDs  of NSPB series were selected at random 
            among 100
            pieces  and were placed in a constant temperature 
           box being kept at constant temperature of 20  $^{\circ}$C 
            .  Each LED was coupled 
             with an optical quartz fiber of 50 $\mu m$ $\phi$  and 430 cm 
              length
             via an optical connector.     \par

             During the whole period of measurement  the three
             LEDs were continuously driven  with pulses of 5.0 volt  in 
             height and of 30 {\it ns} in width at  1 Hz.  The 
             continuous
             driving of the LED at the rate of 1 Hz  for 40 days has supplied
             the LEDs with $\sim$ $3.5 \times 10^{6}$ pulses which
             correspond to driving the same LEDs for about 7 years
             (assuming that the data
             taking period for each year is 70 $ \%\ $ of a year) with 2000 
             pulses a day which is expected to be necessary for  daily
             monitoring in the experiment. \par    

\begin{figure}[p]
\vspace {5.5in}
\hspace{-1.4in}
     \includegraphics{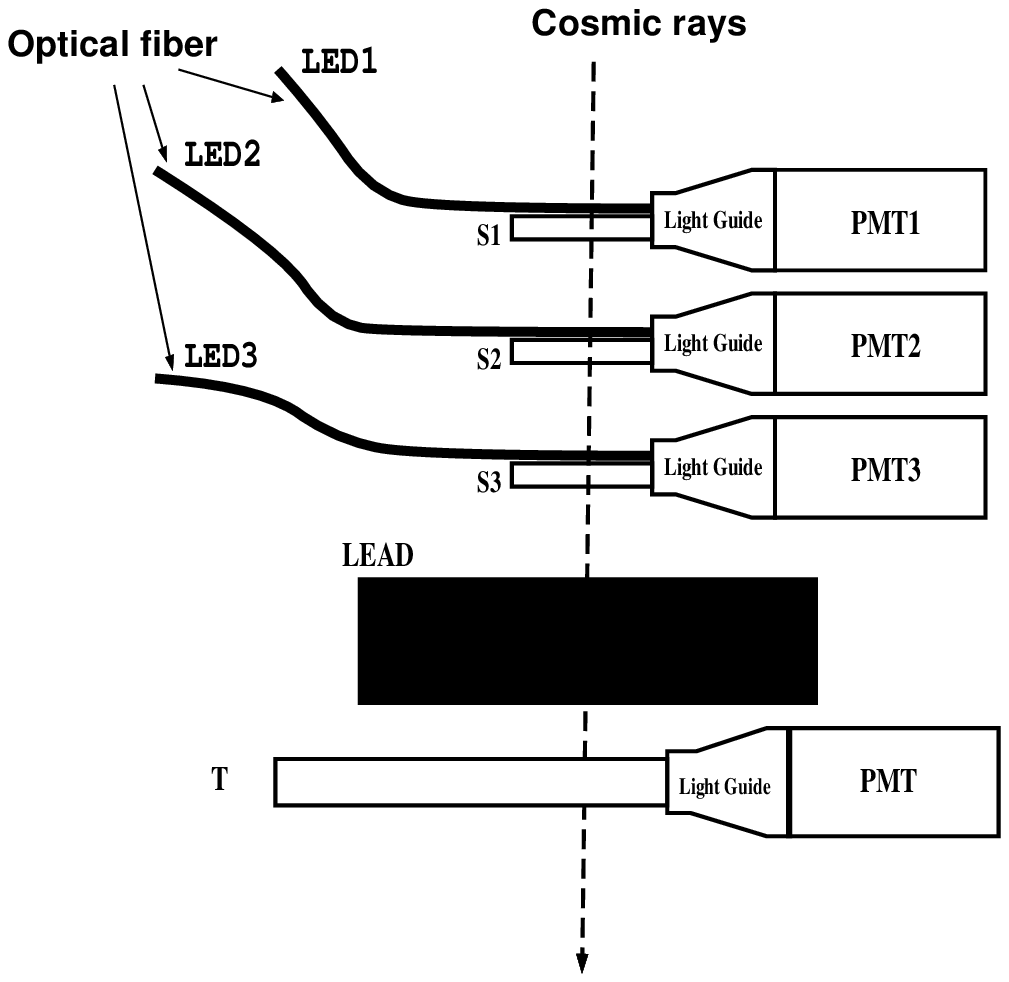} 
\vspace{-2.5in} 
   \caption{  }
\label{fig:setup}
\end{figure}

              Figure \ref{fig:setup}  shows the setup for the 
              measurement 
              of
              the stability
              of the LEDs with respect to time and temperature. The emitted 
              light from
              the three LEDs, LED1, LED2 and LED3 was led to the light guides
              of the three scintillation counters, S1, S2 and S3, respectively
              through 
              the optical fibers. In order to monitor and calibrate the 
              possible drift of the gain of the PMTs used (Hamamatsu H1161),
              energy losses of  cosmic rays in these three 
              scintillation counters ($ 10 cm \times 10 cm \times 0.4 cm $)
              were also measured.  The cosmic rays were hardened by placing
              the lead blocks of 5 {\it cm} thickness between the three 
               scintillation counters and the trigger counter T 
               ($ 30 cm \times 18 cm \times 1.0 cm $). 
               The  whole setup was kept
               at a constant temperature of 25 $^{\circ}$C. \par
                         
               The hardened cosmic rays were triggered by  
               the 
               4-fold coincidence of the signals from S1, S2, S3 and T
               counters with a rate of 
               about 1 Hz. The LED and cosmic ray signals were measured
                simultaneously for about 1 hour every day except 5 days
                during the whole term (40 days) of the measurements. The
                output 
                signals from the PMTs were fed to a CAMAC ADC (LeCroy
                2249W) and digitized data were recorded with a PC9801 
                computer. Pedestals were measured before and after every
                LED/cosmic rays run.    \par
          
            The means of the ADC spectrum for the LED and cosmic ray signals 
            have been obtained by fitting the Gaussian curves to the observed
            spectra.
            The number of photoelectrons $N_{pe}$ for each 
            LED and
            cosmic rays was
            determined by using the relationship  
            \begin{equation}
               N_{pe}=\frac{\mu^{2}}{\sigma^{2}} .  
            \end{equation}
            
         Here, $\mu$ and $\sigma$  are  defined as
         $\mu=\mu_{obs}-\mu_{ped}$  and $\sigma^{2}=\sigma^{2}_{obs}-
       \sigma^{2}_{ped}$ , where $\mu_{ped(obs)}$ and $\sigma_{ped(obs)}$
                  are the mean
                  and the standard deviation of the ADC spectrum for the 
             pedestal (observed LED or
             cosmic ray signal), respectively.    \par

           \subsubsection{Long term stability  of the LED light source}
\begin{figure}[p]
\vspace {6.0in}
\hspace{-1.2in}
     \includegraphics{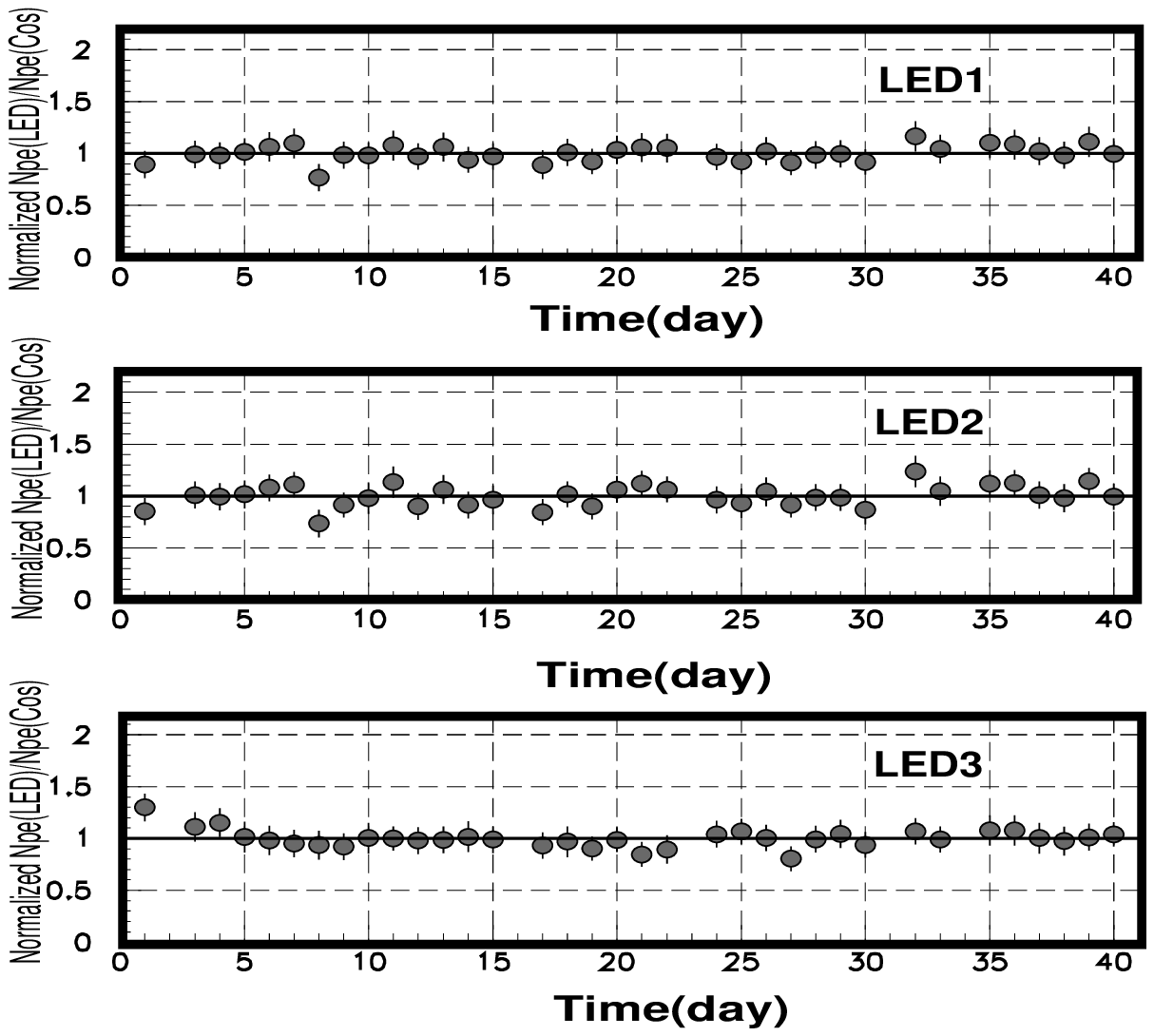}
\vspace{-2.5in} 
   \caption{  }   
\label{fig:longsta2}
\end{figure}

               \hspace {0.2in}  
             The ratios
            of the number of photoelectrons for LEDs to that for cosmic rays, 
           $N_{pe(LED)}$/$N_{pe(Cos)}$ at each day are  normalized to the 
           averaged values 
           over 35 days
           and are plotted in Figure \ref{fig:longsta2} as a 
           function of time.  It should 
           be noted  that the effect of the 
           gain drift of the
           PMTs is  completely cancelled out in this analysis.  
           In the figure the quadratic sum of the    
           statistical errors  and  the Gaussian-fitting ones  
           is shown. \par   
              
           Assuming the linear variation of the light output from the LEDs
           the rate of change of light output has been obtained to be
            (0.096$\pm$0.045)\%\ /day,  ($-0.099\pm$0.059)\%\ /day and
            ($-0.28\pm$0.073)\%\ /day  for LED1, LED2 and LED3,
             respectively. The averaged long term stability for these
            three LEDs is ($-0.094\pm$0.035)\%\ /day . It should be noted
            that in the present measurement one day corresponds to 86400
            pulsing of the LED.                                    
           In the present  study the degree of stability S is  
            defined as
  \begin{equation}
     S=1-\sqrt{\frac{\sum_{i=1}^{n}{({x_{i}}-\bar{x}})^2}{(n-1)}}\frac{1}
      {\bar{x}} , \\
  \end{equation}  
           where $x_{i}$ stands for the i-th measured value and 
           $\bar{x}$=${\sum_{i=1}^{n}}$${x_{i}}$/n  . Using this 
           definition the long term stability of the LEDs
           has been calcuated to be  0.978$\pm$0.007, 0.997$\pm$0.006 and 
           0.965$\pm$0.006 for LED1, LED2 and 
           LED3,  respectively. The average degree of stability  for these
           three LEDs is 0.980$\pm$0.004. \par

           The measured stability together 
           with Fig.8 indicates that  LEDs of NSPB series
           are  stable enough for the 
           long-term use for the   
           present  monitoring system.
  
 \subsubsection{ Temperature dependence of the LED light source}
   \hspace {0.2in}
           Using the same experimental setup as the one  described in 
           subsection 
           3.1.1  and 
           changing  the  temperature in the constant temperature 
           box
           with  a 5 $^{\circ}$C step  from 
           5 $^{\circ}$C to 70 $^{\circ}$C,  the temperature dependence  of 
           the 
           same three LEDs of NSPB series was  measured. It should be
           noted here that the ACC will be kept at  around  25 $^{\circ}$C 
           in the BELLE experiment.
           The LEDs were kept 
           at each 
           temperature for 2  hours to make sure that the LEDs had completely
           reached  the setting temperature. The temperature near the surface 
           of the LEDs was measured continuously with a thermometer  and it
           was observed that within 15 minutes the temperature near the 
           surface of the LED 
           reached  the setting temperature.    \par
 
\begin{figure}[p]
\vspace {6.5in}
\hspace{-1.70in}
   \includegraphics{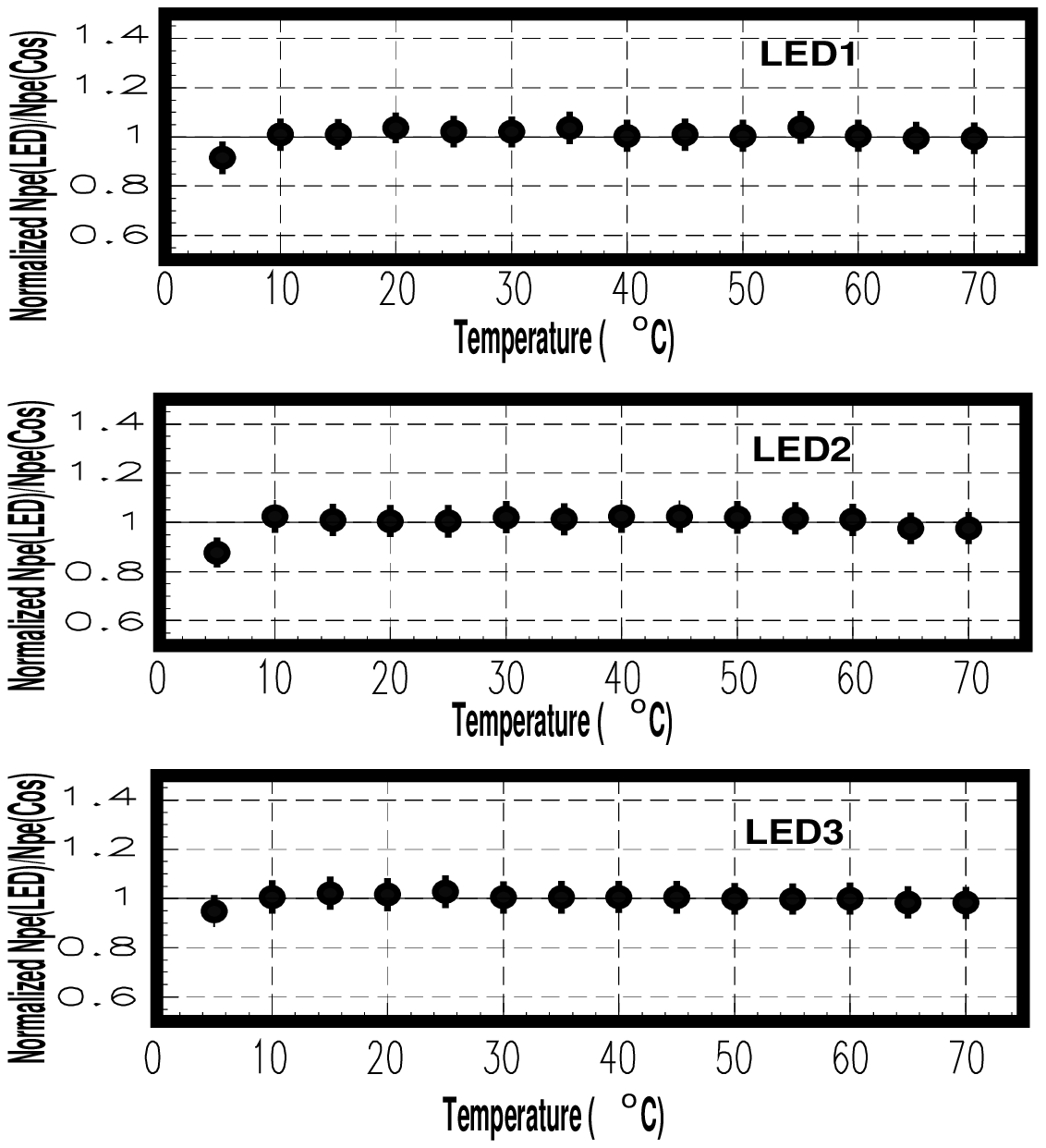}
\vspace{-2in}   
 \caption{  }
\label{fig:temsta2}
\end{figure}

           The ratios
            of the number of photoelectrons for the LEDs to that for  
            cosmic 
            rays, 
           $N_{pe(LED)}$/$N_{pe(Cos)}$ at each temperature  
           are  normalized to the averaged 
           values  over   5 $^{\circ}$C to 70 $^{\circ}$C and  
           are plotted in Figure \ref{fig:temsta2} as a
           function of temperature.  
           Fitting the function exp( $\alpha$T) to these   data in  the
           temperature range from 
            5 $^{\circ}$C to
            70 $^{\circ}$ for each LED   the temperature coefficient $\alpha$
           is obtained to be 
           (0.029$\pm$0.048) \%\ / $^{\circ}$C , ($-0.067\pm$0.057) \%\
             / $^{\circ}$C    and 
             (0.029$\pm$0.074) \%\ / $^{\circ}$C   for LED1, LED2 and LED3, 
            respectively .  The average temperature coefficient for these
            three LEDs is  ($-0.003\pm$0.035) \%\ / $^{\circ}$C . This
             temperature dependence
             is very small compared with  
            the  other  LED light sources used in other experiments.
             The measured 
            temperature dependence indicates that LEDs of NSPB series can
            be reliably used for the ACC monitoring system.
           In Table \ref{tab:tempcomp}  comparison of the performances
           is shown between the present LED light source and the other
           ones. It has been shown that the present LED light source has a 
           much better performance than the others. \par

 \subsection{Performance of the diffuser box}
\begin{figure}[p]
\vspace {7.5in}
\hspace{-2.0in}
    \includegraphics{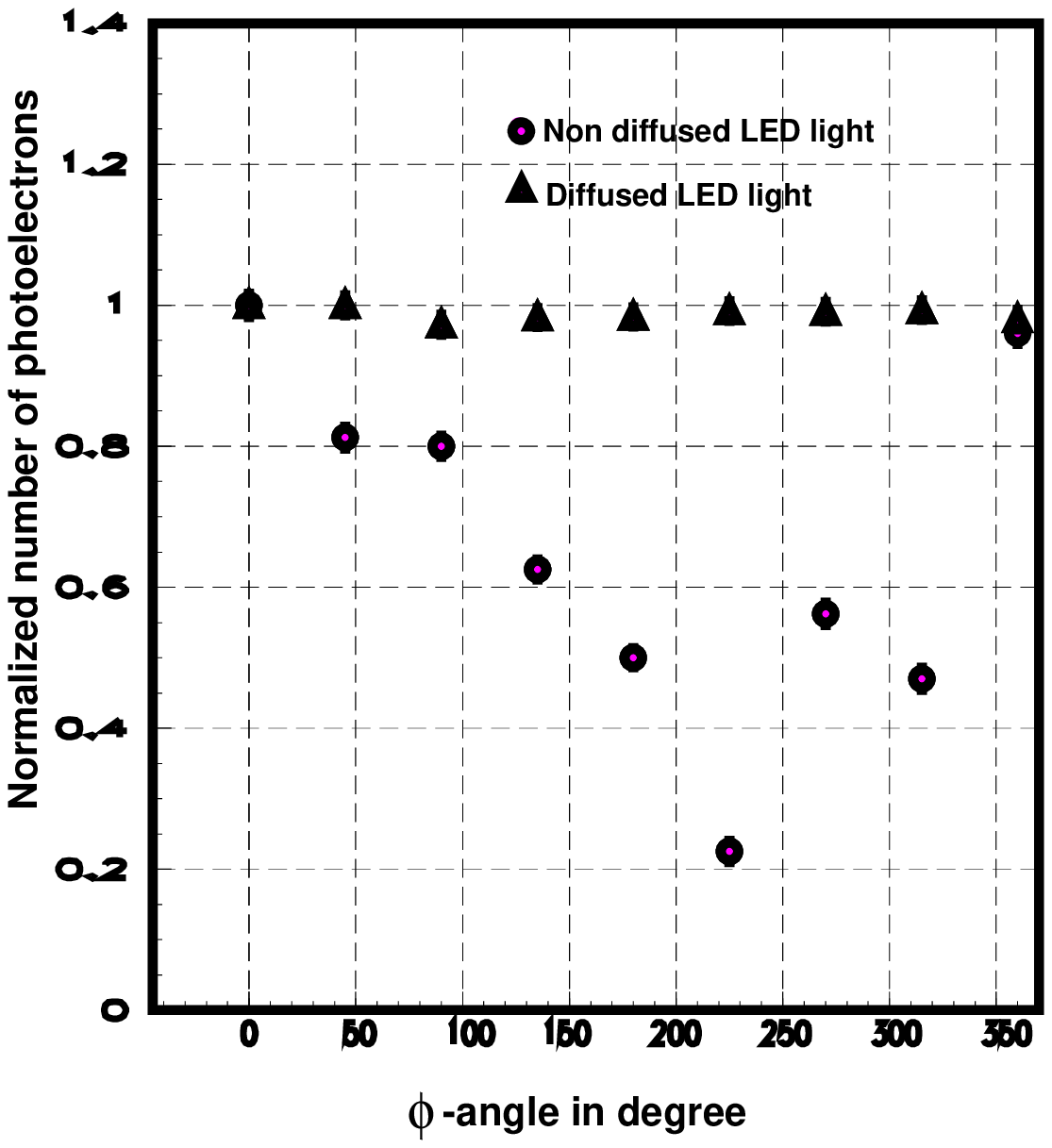}
\vspace{-2.5in}
    \caption{  }
\label{fig:fi_dep}
\end{figure}

       \hspace {0.2in}  
              The  ${\phi}$-angular distributions  of the light intensity
        for the non-diffused
        and diffused LED light   were
        measured.  The light emission axis of the LED was
        defined as the  z-axis for the non-diffused LED light and  the
        light intensity distribution with respect to this axis was measured 
        with a
        45$^{\circ}$  step. 
         The same measurement was performed for the diffused LED light. 
        In this case the center axis of the light out-let hole of the
        diffuser box is defined as the z-axis.
        Figure \ref{fig:fi_dep} 
        shows the  observed  ${\phi}$-angular distributions  of the 
        light intensity for the non-diffiused and 
        diffused LED light. By using the diffuser box  
        the uniformity of  the ${\phi}$-angular distribution  
       of the  light intensity  for the non-diffused LED light was
       improved from   66.0 \%\ to  99.1 \%\ .  This number
       indicates that the diffused light has the complete uniform 
       distribution both in  ${\theta}$  and ${\phi}$ . \par

        It has been  observed that the light intensity of the diffused
        LED light 
        reduces by a 
        factor of $\sim$ 375
        from that of the non-diffused  LED light. 
        The absoluate value of the light intensity 
        through the whole monitoring system with the diffuser box will be 
        discussed  in  subsection 3.4. \par
                      
 \subsection{Performance of the distributor}
       \hspace {0.2in}  
         In order to study  the performance of 
         the distributor network 
        the uniformity of light intensity  among the fibers of each primary
        and secondary fiber bundles was measured.
              The uniformity is defined as
           \begin{equation}
                        Uniformity = 1-\frac{\sigma}{\bar{\mu}},  \\
               \end{equation} 
               where $\bar{\mu}$ is the average  of 
                $N_{pe}$ over 19   fibers of each primary or secondary
                fiber bundle  
                and $\sigma$ is the standard deviation of the distribution 
                of  $N_{pe}$ for 19 fibers. \par
                
        To study the uniformity of  light intensity  from  
        all the secondary fibers  an  LED  was
        driven by 
        the square  pulse of height/width
                 of 4.0 V/30 {\it ns} 
                 at 500 Hz. The LED was directly (without using the diffuser
                 box) coupled to a given primary fiber of a core/clad diameter
                 of 400/500 $\mu m$ with a space gap of 2.0 {\it cm}.
                 The non-diffused LED light from the given primary fiber was
                 transmitted to the secondary fiber bundle.
                 It was confirmed  that the non-uniformity
                 of  light intensity  can be completely neglected on the input
                 surface of the given  primary fiber. The  light output 
                 from each secondary  fiber was  
                 measured with a  line focus PMT
                 of Hamamatsu R329-05S. The output signal of the PMT
           was read with a CAMAC ADC (LeCory 2249W). Using 
           Eq.(1)  the number of photoelectrons  $N_{pe}$ for each secondary
           fiber
           was calculated. \par

\begin{figure}[p]
\vspace {7.5in}
\hspace{-1.0in}
     \includegraphics{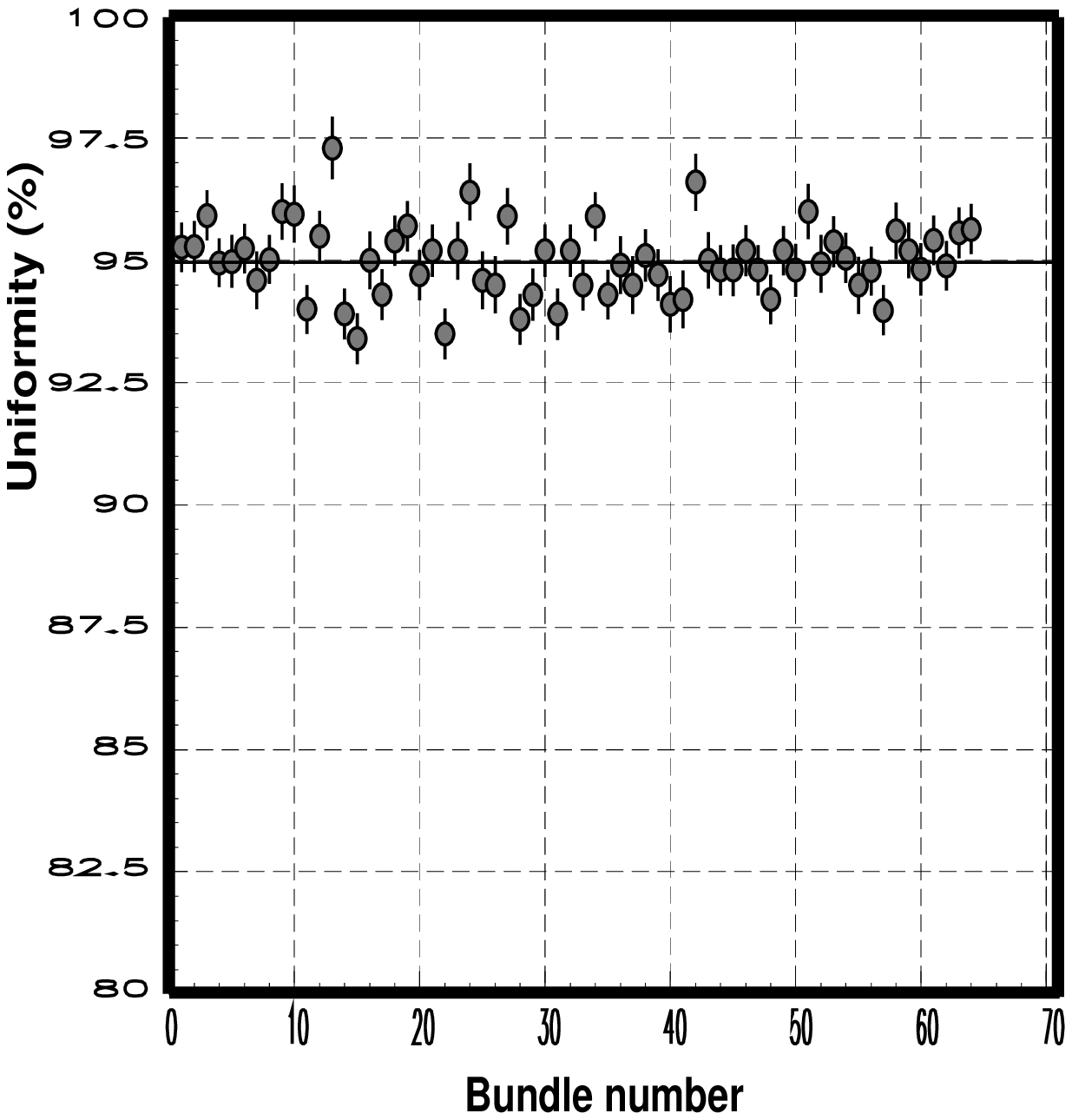} 
\vspace{-2.0in}
    \caption{  }
\label{fig:unif_sec}
\end{figure}

            Measurements of the performance have been made only on the  
            distributors  required for the barrel part ACC.
            The measured uniformity for the 64 secondary fiber bundles is
            shown in Figure \ref{fig:unif_sec}. The average of the uniformity 
            is 94.96$\pm$0.10 \%\ .  \par

          In order to measure the uniformity of light intensity among the
          primary fibers  of each primary fiber bundle the diffused LED
          light was injected to the input surface of each primary fiber bundle.
          The light output from each primary
          fiber was transmitted to a given secondary fiber bundle successively.
          The light
          output  from one of the given secondary 
          fiber  was fed directly to the PMT.
          The  number of
          photoelectrons  was calculated as described 
          above. The uniformity 
          among 19 primary fibers of each primary fiber bundle was calculated
          using Eq.(3).
          \par

          The  averaged uniformity over the 
          four  
          primary fiber bundles has been  measured to be 97.13$\pm$0.32\%\ .
                                               
  \subsection{Light intensity and uniformity of  the whole monitoring system}
\begin{figure}[p]
\vspace {6.0in}
\hspace{-2.5in}
  \includegraphics{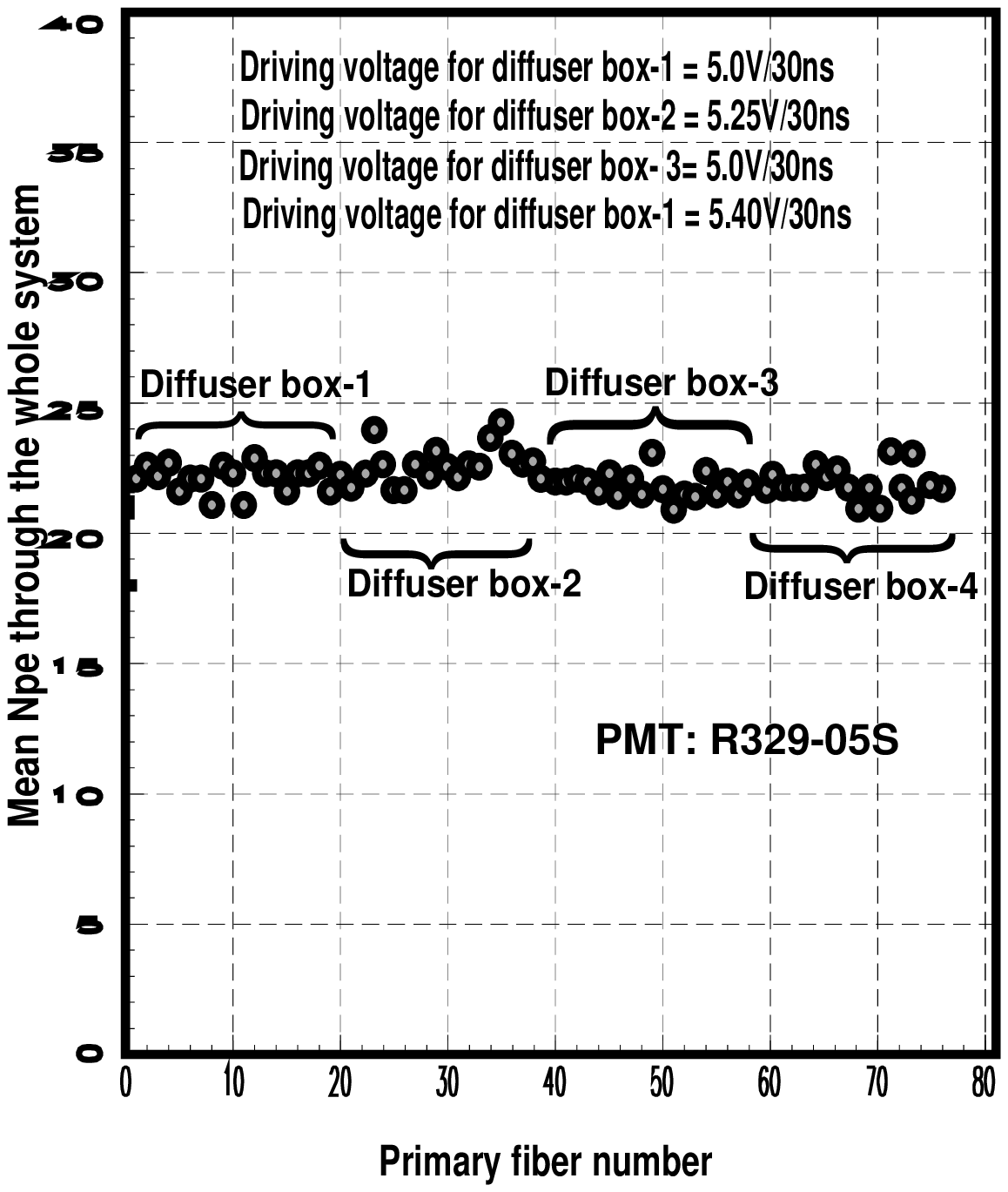}
\vspace{-1.5in}
    \caption{  }
\label{fig:bmen}
\end{figure}

              \hspace {0.2in}
              Measurements were made on the performance of the whole monitoring
              system consisting of the LED light source, the diffuser box and
              the distributor. 
              The six LEDs in the diffuser box were driven with a 
              square pulse of height/width of 5.0 V/30 {\it ns}  at 500
              Hz. It was confirmed that the LED light output does not
              depend on the repetition rate of the LED driving pulse
              in the range from 1 Hz to 10 KHz. The diffused  LED light 
              was transmitted  to the  
              primary fiber  
              bundle through the out-let hole of the diffuser box.
              The light output from each of the 19 primary fibers 
              was transmitted  to a 
              secondary fiber bundle 
              successively. For each primary fiber bundle the light
              intensity  from each secondary fiber of the  secondary fiber
              bundle 
              was measured by using the line focus  PMT  
              R329-05S through the ACC module 
              instead of using a FM-PMT. The same measurement
              was performed for all of the 4 primary fiber bundles.
               Using Eq.(1) the number of photoelectrons  $N_{pe}$ for each
               secondary fiber was calculated. Figure \ref{fig:bmen} shows 
              the observed number of 
              photoelectrons averaged 
              over 19 secondary fibers
              for each primary fiber.
              This figure indicates that by adjusting the LED driving voltage
              for each diffuser box roughly the same light intensity
              can be supplied to each ACC module.  \par

\begin{figure}[p]
\vspace {7.0in}
\hspace{-1.5in}
   \includegraphics{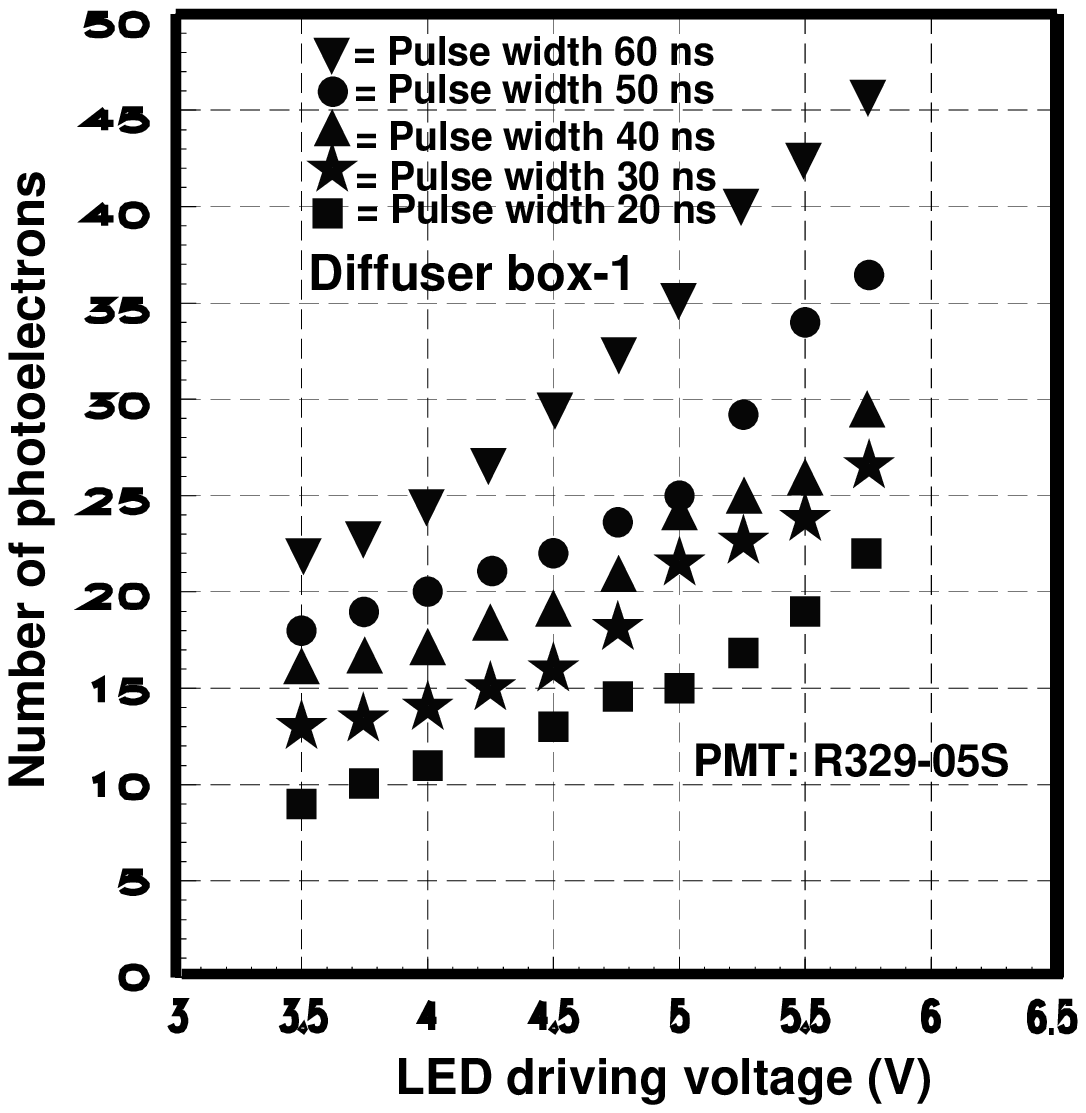}
\vspace{-2.5in}
    \caption{  }
\label{fig:tri}
\end{figure}

              The light intensity  through  the whole
              monitoring system was measured for   the LED driving 
              voltage from 3.5 V to 5.75 V and for the pulse width from 20 
              {\it ns}
              to
              60 {\it ns}.
              In  Figure \ref{fig:tri} the  measured number of photoelectrons
              is shown as a function of the  LED driving voltage.
               This figure shows that
              just controlling the LED driving voltage and/or pulse width
              we can  easily change 
              the number of photoelectrons  in the range  10  $\sim$ 40.
              It should be noted that the  quantum efficiency of  the FM-PMT
              of the ACC is about 
              25 \%\ at 400 {\it nm}  [17] which is nearly the same as that 
              of the  PMT  R329-05S used in the present 
              measurement.
              Therefore the same number of photoelectrons given above, i.e. 
              roughly 10 $\sim$ 40 will be obtained when this monitoring 
              system is employed in the ACC with the FM-PMT. 
              This shows that the developed monitoring system 
              satisfies the requirements of the light intensity for the 
              ACC monitoring  which are described
              in Sec.2. \par 
             
              From  the  uniformity averaged over the 4 primary 
              and
              the 64 secondary fiber bundles which are given in
              Sec.3.3 the   
              non-uniformity of
              light intensity   for the whole system 
              can be obtained using the relation
     \begin{equation}
             (NU)^{2} = (NU_{p})^{2} + (NU_{s})^{2},
     \end{equation}
             where  $ NU_{p}$ and   $NU_{s}$  stand for  the  
             non-uniformity averaged over the 4 primary fiber bundles and  
             that over the 64 secondary 
             fiber bundles, respectively. NU is the   non-uniformity 
             over the whole monitoring  system for the barrel part ACC. \par
                          
             The non-uniformity for the whole monitoring system 
             for the barrel part ACC was then calculated to
             be   5.80$\pm$0.09 \%\ . This means that the present monitoring 
             system has the uniformity of 
              94.20$\pm$0.09 \%\ over  all of the ACC
             modules of the barrel ACC.       \par

 \section{Conclusions}                      
       \hspace {0.2in} 
       We have developed  a new  monitoring system to  be used  to monitor 
       the ACC modules of the BELLE detector at the KEK B-factory.  
        The  employed LED  light source has been observed to have the
        long term stability of  98.0$\pm$0.4\%\  . The
        temperature coefficient of the LED   has
        been measured to be ($-0.003\pm$0.035)\%\ / $^{\circ}$C over
        a temperature range of   5 $^{\circ}$C to 70 $^{\circ}$C.
       In this
       monitoring system  the diffuser box
       is employed  to reduce the intrinsic directional non-uniformity 
       of the LED 
       light intensity.
       The uniformity of light intensity of the developed monitoring 
       system has been measured to be  94.20$\pm$0.09 \%\ . The present
      monitoring system can supply the light intensity corresponding to 
      roughly   10 $\sim$ 40 photoelectrons 
      to each ACC module.
      The performances of the present  monitoring system meets all the
      requirements
      for  monitoring  of all  the ACC modules of the  BELLE detector. 

       \section*{ Acknowledgment}    
       This work was partly supported by Grant-in-Aid for Scientific 
       Research on Priority Areas (Physics of CP violation) from the
       Ministry of Education, Science, and Culture of Japan. We gratefully
       thank all of the members of the ACC subgroup of the BELLE 
       collaboration
       for joining in  discussions, giving comments and advises on this
       work.

%
\newpage
{\bf Figure Captions}  \par    
\vspace{0.1in}
{\bf Fig.1:} Side veiw of the aerogel Charenkov counter system of the BELLE 
detector. The
refractive index {\it n} of aerogel is shown for each type of modules. \par
          
{\bf Fig.2:} A typical single module of the  aerogel Cherenkov counter. The
monitoring light is fed to the ACC module from the light distributor   
through the fiber connector. \par    

{\bf Fig.3:} The structural view of the diffuser box(not in scale). Emitted 
 light from 6 LEDs is reflected and diffused by the reflector BaSO$_{4}$ on
 the inner surface. The diffused LED light is transmitted to the primary
 fiber bundle through an out-let hole on the light injection surface. \par    

{\bf Fig.4:} One complete set of the light distributor network for the 
BELLE ACC monitoring  system.  Each primary quartz fiber has a length of
 10 {\it m} . The secondary quartz fibers has a length of 0.5 $\sim$ 2.0 
{\it m}. \par

{\bf Fig.5:}The structural view of the connection part of the diffuser box with
the primary fiber bundle (not in scale). The mode scrambler is several 
circular turns of each primary fiber having a radius of 1.75 {\it cm}. \par

{\bf Fig.6:}The structural view of the connection part of the primary fiber 
with the secondary fiber bundle (not in scale). To get the uniform LED 
light intensity
on the input surface of the secondary fiber bundle 3.6 {\it mm} spacing is used
between the primary fiber and the secondary fiber bundle. \par

{\bf Fig.7:}Experimental setup for the stability measurement of the LEDs of
NSPB series (not in scale). In order to monitor and calibrate the possible
gain drift of the PMTs , the energy losses of cosmic rays in S1, S2 and S3
was measured. Light from three LEDs was fed to the light guide of the
scintilation counters with three optical fibers. \par

{\bf Fig.8:}Long term stability of the LEDs of NSPB series.  The ratios
of the number of photoelectrons for the LED to that for cosmic rays, 
$N_{pe(LED)}$/$N_{pe(Cos)}$  normalized to the averaged values  over 35 days
are plotted as a function of time. Only statistical and Gaussian fitting 
errors are shown. \par

{\bf Fig.9:}Temperature dependence  of the  light intensity of the LEDs of
NSPB series.
The ratios
of the number of plotoelectrons for the LED to that for cosmic rays, 
$N_{pe(LED)}$/$N_{pe(Cos)}$ are normalized to the averaged values 
over  5 $^{\circ}$C to 70 $^{\circ}$C.  Only statistical and Gaussian fitting 
errors are shown.  \par

{\bf Fig.10:} $\phi$-angle distribution  of the non-diffused  and  
diffiused  LED light. The $\phi$-angle at which the intensity is maximum is
defined as zero degree. All the data
points are normalized to the number of photoelectrons obtained at zero degree.
Only  statistical errors are shown in the figure. \par

{\bf Fig.11:}Uniformity of  light intensity of the 64 secondary fiber bundles.
Each data point shows
the uniformity among the 19 secondary fibers of each secondary fiber 
bundle.  Only statistical and Gaussian fitting errors are shown. \par

{\bf Fig.12:}The light yield  through the whole
monitoring system for each primary fiber.  Each data point
shows  the  number of photoelectrons averaged over 19 secondary
fibers of  the given secondary fiber bundle.  The error bars are within
the circle of the data points. \par

{\bf Fig.13:} LED driving voltage vs the light intensity of the LED of NSPB
series through the whole
 monitoring system. The 6 LEDs  in the  diffuser box were 
driven with
square pulses of height from 3.5 V to 5.5 V and of width from 30 {\it ns} to 
60 {\it ns}.
Only  statistical errors are shown in the figure. The error bars are within 
the circle of the data points. \par 


\newpage
\begin{table}[h]
\caption{Statistics  of the light distributor network for the
BELLE ACC monitoring system.}
\label{tab:dist}
\begin{center}
\begin{tabular} {l l l}  \\
\hline   
\hline
Parameters &  Primary fiber  & Secondary fiber   \\
           &  (barrel+endacp) & (barrel+endcap)  \\
\hline
Number of sets & (4+1)   &  (4+1)  \\  
Number of used bundles/set    &  (1+1)    & (15+15)  \\
Number of spare bundles  & (0+0)  &  (4+2)  \\
Total number of bundles  &  (4+1)   &  (64+17)  \\
Number of used fibers/bundle & (15+15)  & (16+16)  \\
Number of spare fibers/bundle & (4+4) & (3+3) \\
Total number of fibers/set  & (19+19) & (285+285) \\
Total number of used fibers  & (60+15) & (1024+240) \\
Total number of  fibers  & (76+19) & (1216+323) \\
Diameter of a fiber bundle & 3.0 {\it mm}  &  600 $\mu m $  \\
Length/diameter of each fiber & 10 {\it m} , 400/500 $\mu m$ & 0.5 $\sim$ 2.0 {\it m},110/125 $\mu m$   \\ 
\hline
\hline
\end{tabular}
\end{center}
\end{table}
\vspace{2in}
\begin{table}[h]
\caption{Comparison of the temperature dependence between the present LED 
of NSPB series
and some other types of LEDs.}
\label{tab:tempcomp}
\begin{center}
\begin{tabular} {l l l l}  \\
\hline
\hline  
Experiment &LED type/Material & Temperature  &Temperature  \\
            &   &  range ( $^{\circ}$C)  & coefficient( \%\ / $^{\circ}$C)  \\
\hline 
Present study & blue/GaN & 5 $\sim$ 70 &  $-0.003\pm$0.035 \\
WA98[14]    & blue/SiC & 10 $\sim$ 60 & $-$1.1  \\
PHOENICS[15] &  blue &   & $-1.0\pm$0.05 \\
CsI(BELLE)[16]  & blue  & 23 - 31        &  $-0.7$          \\
\hline
\hline			
\end{tabular}
\end{center}
\end{table}
%
\input{epsf.sty}
\clearpage
%
\clearpage
%
\clearpage
%
\clearpage
%
\clearpage
%
\clearpage
\clearpage
%
\clearpage
%
\clearpage
%
\clearpage

\clearpage
%
\clearpage
%
\clearpage

\newpage
\begin{thebibliography}{99}
        \bibitem{kn:gnus} T. Iijima et al., 
                {\em Nucl. Instrum. Meth. {\bf A379}(1996)457.}
        \bibitem{kn:gnus} M. Hazumi, 
                {\em Nucl. Phys. B(Proc. suppl.) {\bf 59 }(1997)61.} \\
                 { T Nozaki,} 
                 {\em Nucl. Phys. B(Proc. suppl.) {\bf 50 }(1996)288.} 
        \bibitem{kn:gnus} BELLE Collaboration,
                {\em ``Technical Design Report'', KEK procedings 95-1.}        
        \bibitem{kn:gnus} I. Adachi  et al.,   
                  {\em Nucl. Instrum. Meth. {\bf A355 }(1995)390.}
        \bibitem{kn:gnus} T. Iijima et al., 
                 {\em Nucl. Instrum. Meth. {\bf A387}(1997)64. }
        \bibitem{kn:gnus} R. Enomoto  et al.,
                  {\em Nucl. Instrum. Meth. {\bf A332 }(1993)129.}             
        \bibitem{kn:gnus} F. Bonutti et al.,
                 {\em Nucl. Instrum. Meth. {\bf A337 }(1993)165.}\\
                   {  T. Sumiyoshi et al.,}
                  {\em Nucl. Instrum. Meth. {\bf A271 }(1988)432.}
        \bibitem{kn:gnus} D. Autiero et al.,
                  {\em Nucl. Instrum. Meth. {\bf A372 }(1996)556.} \\ 
                   { Janusz Zabierowski,} 
                   {\em Nucl. Instrum. Meth. {\bf A388 }(1994)577.} \\  
                   { J. Berger et al.,}
                   {\em Nucl. Instrum. Meth. {\bf A279 }(1989).} \\
                   { Ronald J. Madaras, Barrie Pardoe and Ruben Pecyner,}
                    {\em Nucl. Instrum. Meth. {\bf 160 }(1979)263.} \\  
                     {  G. Anton, K. Buchler and M. Kuckes ,}
                  {\em Nucl. Instrum. Meth. {\bf A274 }(1989)222.}     
        \bibitem{kn:gnus} OPAL Collaboration,
                    {\em Nucl. Instrum. Meth. {\bf A305 }(1991)275.}   
        \bibitem{kn:gnus} See for example, S. Bianco  et al.,
                  {\em Nucl. Instrum. Meth. {\bf A305 }(1991)48.}   
        \bibitem{kn:gnus} Catalogue of 
                  {\em NICHIA Chemical Industries Ltd., Japan, Cat. No. 9603.} 
        \bibitem{kn:gnus} S. Nakamura, M. Senoh, N. Iwasa and 
                    S. Nagahama,
                    {\em Jpn. J. Appl. Phys. {\bf 34, Part2, }(1995)L797.} \\ 
                   {S. Nakamura, M. Senoh, N. Iwasa, S. Nagahama, T. Yamada
                     and T. Mukai,}
                  {\em Jpn. J. Appl. Phys. {\bf 34, Part2, }(1995)L1332.}   
        \bibitem{kn:gnus} Private communication with
                   {\em NICHIA Chemical Industries Ltd., Japan.} 
        \bibitem{kn:gnus} T. Peitzmann  et al., 
                  {\em Nucl. Instrum. Meth. {\bf A376 }(1996)368.}  
        \bibitem{kn:gnus}  G. Anton, K. Buchler and M. Kuckes ,
                  {\em Nucl. Instrum. Meth. {\bf A274 }(1989)222.}     
        \bibitem{kn:gnus} BELLE Collaboration,   
               {\em  ``BELLE Progress Report``  1995 April-1996 March, p. 63.}
        \bibitem{kn:gnus} PMT Catalogue of 
                  {\em Hamamatsu Photonics K.K., TPMH 1110E01, Oct., 1995.}
\end{thebibliography}
\end{document}